\documentclass[preprint,3p,twocolumn,times]{elsarticle}

\usepackage{amssymb}

\usepackage[T1]{fontenc}
\usepackage{amsmath}
\DeclareMathAlphabet{\altmathcal}{OMS}{cmsy}{m}{n}
\usepackage{lipsum}
\usepackage[dvipsnames]{xcolor}
\usepackage[normalem]{ulem}
\usepackage{soul}

\makeatletter
\def\cat@comma@active{\catcode`\,12}%
\makeatother

\journal{Results in Engineering}

\begin{document}

\begin{frontmatter}



\title{Active grid turbulence anomalies through the lens of physics informed neural networks}

\author[aff1]{Sof\'ia Angriman}
\author[aff5,aff6]{Sarah E. Smith}
\author[aff2]{Patricio Clark di Leoni}
\author[aff3,aff4]{Pablo J. Cobelli}
\author[aff3,aff4]{Pablo D. Mininni}
\author[aff6,aff7]{Mart\'in Obligado}
\affiliation[aff1]{organization={Physics of Fluids Group, Max Planck Center for Complex Fluid Dynamics, and J. M. Burgers Centre for Fluid Dynamics, University of Twente, P.O. Box 217, 7500AE Enschede}, country={The Netherlands}}

\affiliation[aff5]{organization={Department of Mechanical and Materials Engineering, Portland State University, Portland, 97207, OR}, country={USA}}

\affiliation[aff6]{organization={Universit\'e Grenoble Alpes, CNRS, Grenoble-INP, LEGI, Grenoble, F-38000}, country={France}}
             
\affiliation[aff2]{organization={Departmento de Ingenier\'ia, Universidad de San Andr\'es, Buenos Aires}, country={Argentina}}
                                
\affiliation[aff3]{organization={Universidad de Buenos Aires, Facultad de Ciencias Exactas y Naturales, Departamento de F\'isica, Ciudad Universitaria, 1428 Buenos Aires}, country={Argentina}}

\affiliation[aff4]{organization={CONICET-Universidad de Buenos Aires, Instituto de F\'isica Interdisciplinaria y Aplicada (INFINA), Ciudad Universitaria, 1428 Buenos Aires}, country={Argentina}}

\affiliation[aff7]{organization={Univ. Lille, CNRS, ONERA, Arts et Metiers Institute of Technology, Centrale Lille, UMR 9014-LMFL-Laboratoire de M\'ecanique des Fluides de Lille - Kamp\'e de F\'eriet, Lille, F-59000}, country={France}}

\begin{abstract}
Active grids operated with random protocols are a standard way to generate large Reynolds number turbulence in wind and water tunnels. But anomalies in the decay and third-order scaling of active-grid turbulence have been reported. We combine Laser Doppler Velocimetry and hot-wire anemometry measurements in a wind tunnel, with machine learning techniques and numerical simulations, to gain further understanding on the reasons behind these anomalies. Numerical simulations that incorporate the statistical anomalies observed in the experimental velocity field near the active grid can reproduce the experimental anomalies observed later in the
decay. The results indicate that anomalies in experiments near the active grid introduce correlations in the flow that can persist for long times.
\end{abstract}







\end{frontmatter}


\section{Introduction}

Since the first active grid was proposed in the early 1990s \cite{makita1991active}, active grids have become a standard instrument for generating bespoke turbulent flows in wind and water tunnels \cite{Mydlarski2017}. These devices are composed of rotating blades that can be operated independently and, therefore, can be used to tune the large scales of the flow, allowing for the tailoring of inhomogeneous velocity profiles \cite{Hearst2015,hearst2017tailoring} and even unsteady conditions, such as gusts and velocity steps \cite{neuhaus2021exploring,wester2022design}. These properties also make them of interest for studying turbulent wakes under different inflow conditions, allowing for the simulation of field conditions for scaled wind turbine rotors~\cite{scott2020wind,hamilton2012statistical,gambuzza2021effects}.

One of the most widespread uses of active grids is to generate moderate-to-high Reynolds numbers in a wind tunnel with large values of turbulent intensity. By operating the blades in various random protocols, the lack of a characteristic time scale at the turbulence generation results in integral time and length scales that are on the order of the wind tunnel's lateral size instead of the mesh size, as is the case with static grids~\cite{makita1991active}. This strategy allows to increase the separation of scales, reaching Reynolds numbers based on the Taylor microscale as large as $R_\lambda \sim 1500$ in standard wind tunnel facilities~\cite{larssen2011generation}. Moreover, for variable density facilities, values as large as $R_\lambda \sim 5000$ have been reported \cite{bodenschatz2014variable,sinhuber2017dissipative}. Furthermore, random protocols still generate turbulent flows that are close to homogeneous and isotropic turbulence (HIT) conditions~\cite{mydlarski1996onset,Kuchler2023} (meaning that the flow is as close to HIT as for regular static grids).  Consequently, active grids have been used to study fundamental turbulence~\cite{Mora2019,sinhuber2015decay,zheng2023unsteady} and even two-phase flows~\cite{Obligado2014,sumbekova2017preferential,prakash2016energy,ferran2023experimental}.

While active grids operated with random protocols have been extensively used in several studies, many open questions remain concerning the properties of the turbulent flow they generate. For instance, kinetic energy has been found, in some cases, to decay in space and time following a power law with exponents different from those reported in static-grid-generated turbulence~\cite{kang2003decaying}. Also, anomalous behaviour has been reported for the compensated longitudinal averaged structure function $S_3(\ell)/\left( \varepsilon \ell \right)$, where $\varepsilon$ is the averaged turbulent kinetic energy dissipation rate, $\ell$ the spatial increment, and $S_3(\ell)$ is defined as
\begin{equation}\label{eqS3}
	S_3(\ell) = \langle [u^\prime(x + \ell) - u^\prime(x)]^3\rangle,
\end{equation}
with $u^\prime(x,t)$ the fluctuating streamwise velocity. While for homogeneous isotropic turbulence, within the Kolmogorov phenomenology, the value of $S_3(\ell)/\left( \varepsilon \ell \right)$ should be equal or lower than 0.8 \cite{Frisch1995}, experiments in active grids have reported results above this number~\cite{zheng2021turbulent} for streamwise distances $x$ as large as $x/M=30$ (with $M$ the mesh size of the active grid). For larger streamwise distances ($x/M>80$), this anomaly is no longer observed~\cite{mydlarski1996onset}. Moreover, it has been reported that at $x/M=75$ the far field already presents two-point statistics independent of  $R_\lambda$~\cite{Kuchler2023,kuchler2021measurements}. Note that for passive grids, turbulence is expected to be fully developed in the range $20<x/M<50$~\cite{batchelor1948decay,krogstad2010grid}, whilst for active grids such distance is expected to be much shorter~\cite{larssen2005large,larssen2011generation}. As a result, statistical anomalies in active-grid-generated flows seem to last for longer streamwise distances than for their static counterparts, before reaching a universal behaviour close to the one predicted within the Richardson-Kolmogorov phenomenology. While some of the reported anomalies are probably related to the persistence of inhomogeneities and/or anisotropy, no clear explanation has been provided to assess the anomaly in $S_3(\ell)$ in the near field of the wake.

Another issue that arises when comparing active-grid turbulence generated with random protocols to other turbulent flows is the difficulty in defining time and length scales that would allow to compare different statistics among them. For instance, the autocorrelation function of velocity also presents anomalies, as in some cases it never crosses zero, making it difficult to define an integral length scale~\cite{puga2017normalized,mora2020estimating}. This issue arises when purely random protocols are used, implying that the forcing imposed by the grid is statistically unsteady. The lack of clearly defined length scales also complicates the delineation of the production range in the near field of the turbulent flow. Moreover, the very high values of turbulence intensity near the grid (which can reach 50\% or even more) make it challenging for some standard collection techniques, such as hot-wire anemometry, to properly quantify the flow. Indeed, the use of Taylor's hypothesis in these flows presents certain limitations~\cite{shet2017eulerian}. Furthermore, the large scale separation in these flows makes them also extremely difficult to be characterised via standard optical techniques.

The present work aims at gaining further understanding about the decay of active-grid-generated turbulence using random protocols. Given the significant problems related to characterising this flow experimentally, we turn as well to numerical simulations.
In practice, Laser Doppler Velocimetry (LDV) is initially used to characterise the turbulent flow at $x/M=3$. This technique results in a resolved two-dimensional (2D) map of velocity (including the average value and higher moments) in the plane perpendicular to the freestream velocity.
Direct numerical simulations (DNSs) are then used to understand in more detail the flow physics, as they provide access to the full velocity field, with all dynamical scales properly resolved, and hence they allow computation of gradients and correlations. Note that turbulence in wind tunnels is often compared with simulations of homogeneous and isotropic turbulence. In particular, the flow at different distances from the grid is comparable to the numerical evolution of a flow in the absence of forcing (i.e., {\it freely decaying}), where the initial condition is usually consistent with a fully developed turbulent state.
In our case, in order to generate proper initial conditions for 3D DNSs that can resemble the near-active-grid flow in a statistical sense, we use a protocol which combines Physics-Informed Neural Networks (PINNs) \cite{Raissi2019} with a data assimilation technique known as {\it nudging} \citep{Clark2018, Clark2020}. The combination of the two has been shown to be successful at generating turbulence-compatible velocity fields with fixed given statistical moments \citep{Angriman2023}. This approach allows us to create a 3D box that is evolved in time using the DNS code, fully capturing and characterising  the decay of turbulence. For the sake of comparison, we also tested synthetic initial conditions that correspond to HIT. Hot-wire anemometry (HWA), performed at $x/M=30$, finally allows us to test small-scale information (such as $S_3(\ell)$) in the wind tunnel experiment, aiming at verifying the anomalies previously reported in active-grid-generated turbulence, and matching the timescales between DNSs and experiments. In summary, the methodology used in this study is the following: (1) We use LDV measurements to characterize the flow near the active grid. (2) We used PINNs and {\it nudging} to assimilate the statistical information of these measurements in DNSs. (3) We compare the DNSs time evolution with HWA wind tunnel measurements downstream.

\section{Experimental set up and measurements}
\begin{figure*}[t!]
\centering
\includegraphics[width=1.0\textwidth]{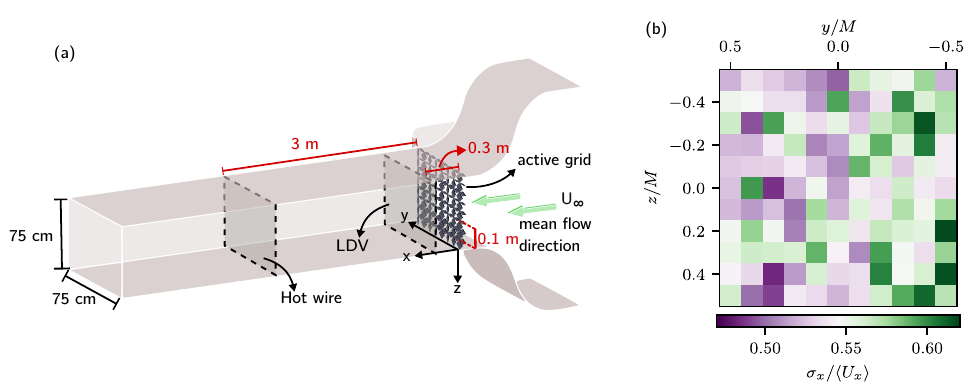}
\hfill
\includegraphics[width=1.0\textwidth]{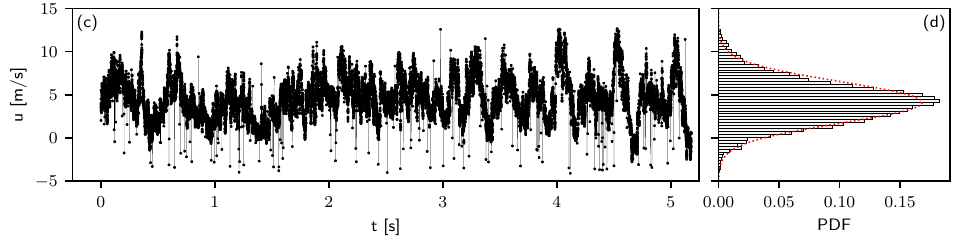}
\hfill
\includegraphics[width=1.0\textwidth]{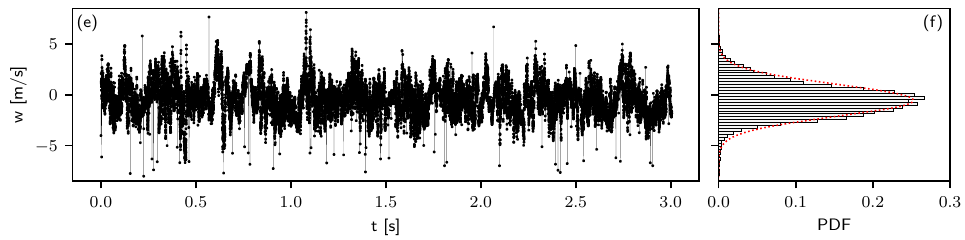}
\caption{(a) Schematic of the wind tunnel in LEGI where experiments were carried out (representation not to scale), showing the position where the laser Doppler velocimetry (LDV) and hot-wire anemometry (HWA) measurements were taken. (b) Turbulence intensity heatmap from LDV measurements. Panels (c) and (e) show raw LDV velocity measurements of the streamwise, $u$, and spanwise, $w$, velocity components respectively, while (d) and (f) show the probability density functions (PDFs) of each signal.}
\label{fig:WT}
\end{figure*}

We carried out experiments in the Lespinard wind tunnel at LEGI, Grenoble, France. This closed-loop wind tunnel has a test section of $4$~m long and a cross section of $0.75 \times 0.75$~m$^2$. Turbulence was generated by means of an active grid (denoted herein as AG), comprised of 16 rotating axes, eight horizontal and eight vertical, each mounted with coplanar square blades. Both the grid mesh size and the blades have a size equal to $M = 10$~cm. We used the grid in `triple-random mode', i.e., the rotation rates and directions were varied randomly in time to random values. As stated in the previous section, this protocol is expected to generate a near-HIT flow (see more details in \citep{Obligado2014, Mora2019}). In this way, we generate a turbulent flow with a mean flow velocity $U_\infty = 4.6$~m/s in the streamwise direction ($\hat{x}$), measured in the region where turbulence is developed. Panel (a) of Fig.~\ref{fig:WT} shows a schematic representation of the setup, depicting the coordinate system, the wind direction, and the active grid. 

\begin{figure*}
\centering
\includegraphics[width=1.0\textwidth]{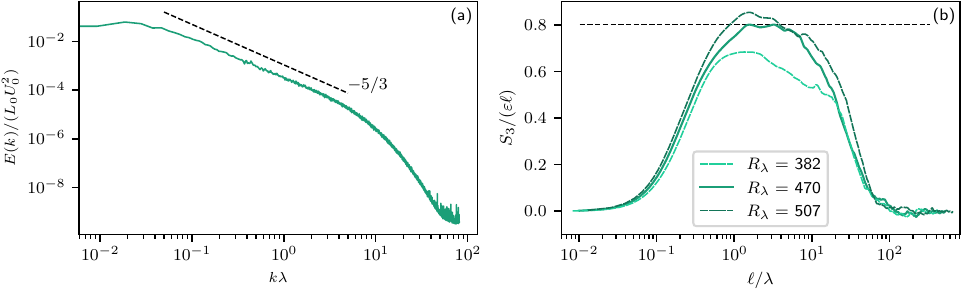}
	\caption{Hot-wire (HW) measurements taken downstream of the wind-tunnel at $x_{HW}/M=30$, for $R_\lambda = 470$. (a) One dimensional kinetic energy spectrum $E(k)$, computed from the streamwise velocity $u$, non-dimensionalised by a large scale velocity $U_0$ and a large-scale length scale $L_0$. (b) Third order velocity structure function, also computed from $u$, and normalised by $\varepsilon \ell$. The black dashed horizontal line indicates $4/5$. In panel (b) we also show the structure function corresponding to two other Reynolds numbers.
	}
\label{fig:HW}
\end{figure*}

\subsection{The flow near the active grid: Laser Doppler Velocimetry measurements}

We performed measurements at $x_{LDV} = 0.3$~m ($3M$) downstream of the grid with an LDV system. This position corresponds to the region closest to the AG which is optically accessible. In particular, we used the closed measurement system LaserExplorer (Dantec Dynamics), which provides access to two components of the velocity field ${\bf u} = u \hat{{\bf x}} + v \hat{{\bf y}} + w \hat{{\bf z}}$. We measured the velocity components in the streamwise direction $u$ and in the vertical direction $w$, in a 2D plane parallel to the grid, and centred about the geometrical centre of the tunnel section, covering a total area of $10\times 10$~cm$^2$ (i.e., one entire mesh size in each direction), with a separation of $1$~cm in each direction. 
For the measurements that we performed, the velocity resolution was of $1\times 10^{-3}$~m/s.

Figure~\ref{fig:WT}(b) shows a heat map of the turbulence intensity, computed from the r.m.s. value $\sigma_x$ of streamwise component of the velocity fluctuations $u^\prime$, normalised by the mean velocity $\langle U_x \rangle$ in the region spanned by our measurements. The fluctuations are computed by averaging in time the signal at each of the measurement points, while the value of $\langle U_x \rangle$ corresponds to the ensemble average of the mean velocity of each measurement point. We observe a high level of turbulence intensity, which ranges from around $48$\% up to more than $60$\% in some points, with no clear structure or identifiable pattern.
Note that there is no temporal correlation between the measurements in each of the grid points, as the whole LDV system has to be repositioned each time the observed point was changed. 
However, we verified that the observed behaviour is systematic and repeatable (in a statistical sense). In consequence, it is probably related to small differences in the shafts and motor responses due to small design irregularities and the ageing of the system.
The high level of fluctuations can also be observed in panel (c) of Fig.~\ref{fig:WT}, which shows the time signal $u(t)$ in a given measurement point. Interestingly, the fluctuations are high enough to observe flow reversal (i.e., negative velocities). We observe a similar behaviour for the spanwise velocity $w$, whose time evolution is shown in Fig.~\ref{fig:WT}(e), where a high level of fluctuations is also present (albeit with a mean value close to zero, as expected). Panels (d) and (f), also in Fig.~\ref{fig:WT}, show the probability density functions (PDFs) corresponding to the temporal velocity signals $u$ and $w$, respectively. Normal distributions with the same mean and standard deviation as the data are shown as references. We observe that the the PDFs are non Gaussian, and that they present an asymmetry, and we observe a similar trend in the other points. We quantify this asymmetry by estimating the centralised third-order moment, $s$, of the streamwise velocity component $u$ at a point $i$ on the measurement grid,
\begin{equation}
	s_i = \Big [\Big \langle \big(u_i - \langle u_i\rangle\big)^3 \Big\rangle \Big ]^{1/3},
\end{equation}
where $\langle \cdot \rangle$ indicates a time average.
Note we take the cubic root (as compared to the standard definition of the moment of a distribution) as this presents some numerical benefits in the setup of the simulations, as we will discuss later. Then, the overall ``skewness'' \footnote{We use here the term skewness in a loose way as a synonym of third-order centralised moment.} $s_{LDV}$ over the entire measurement region is computed by averaging $s_i$ over all of the sampling points, which yields
	\begin{equation}
	\frac{s_{LDV}}{\langle \sigma_x \rangle} = 0.51 \pm 0.15,
	\end{equation}
where $\langle \sigma_x \rangle$ represents the average streamwise velocity fluctuations over the domain. 
Across the entire region $s_i$ presents variations compatible with those observed for the velocity fluctuations (see panel (b) in Fig.~\ref{fig:WT}), and is consistently different from zero, i.e., the deviation from Gaussianity that we observe near the grid is systematic.
We highlight that other measurement techniques, such as hot-wire anemometry (see next section), are not capable of capturing the flow reversal events, so the use of the LDV technique is key to characterise the flow in the region near the AG. 

\subsection{Fully developed turbulence region: Hot-Wire measurements}

Under the same flow conditions used for the LDV measurements, we performed measurements with a hot wire (HW) probe, positioned at $x_{HW} = 3.0$~m ($30M$) downstream of the active grid, and at the centre of the tunnel. We used a Dantec Dynamics 55P01 probe, with a constant temperature anemometer Dantec StreamLine. 
We collected data for $180$~s with a sampling frequency of $50$~kHz.
As discussed in the previous section, the position of the HW corresponds to the typical location where turbulence is expected to be fully developed, at a distance roughly equal to $30M$ but some anomalous behaviour, particularly in terms of the third-order longitudinal structure function, may be expected. Since at this position the turbulence intensity is approximately $17.6$\%, we make use of Taylor's frozen-turbulence hypothesis to reinterpret the measured time signal $u(t)$ as a space-dependent signal $u(x)$. Conversely, we can interpret the fixed position of the hot-wire probe as a fixed time in the evolution of the free decay of the turbulence generated at the grid. 
That is to say, if we follow a turbulent patch generated at the inlet of the tunnel as it is advected by the mean wind with velocity $U_\infty$, its time of flight from the grid to the hot-wire will be $t = x_{HW}/U_\infty$.

In Fig.~\ref{fig:HW}(a) we show the kinetic energy spectrum $E(k)$, made non-dimensional by normalising it by $L_0 U_0^2$, as a function of the wave-number $k$.
Here, $U_0$ is estimated from the measurements done with the LDV system as $U_0\equiv \langle \sigma_x\rangle$. The integral length-scale is computed from the zero-crossings of the longitudinal velocity fluctuations $u^\prime$ \citep{mora2020estimating}, as the scale for which a low-pass filter will result in zero crossings that are decorrelated in time (or space, given that the Taylor hypothesis is used).
The spectrum displays the typical behaviour expected for a turbulent flow, namely nearly two decades with a power-law-compatible range, with an exponent close to the predicted $-5/3$, followed by a dissipative range at the smallest scales (largest wavenumbers). To compute the energy spectrum we take only the fluctuating part of the signal, and we use Welch's method with an overlap of $25\%$ and a Hanning window. The use of Welch's method helps in reducing the noise, especially at the highest frequencies; we have verified that it yields similar results for our dataset as computing $E(k)$ using other methods, such as the Fourier transform of the velocity auto-correlation function. From this spectrum we can estimate the energy dissipation rate, $\varepsilon$, by means of the relation $\varepsilon = \int 15 \nu k^2 E(k)~ dk$, where $\nu$ is the fluid kinematic viscosity. At this distance from the grid the Taylor-scale Reynolds number then results $R_\lambda =  \sigma_x \lambda /\nu \approx 470$, where the Taylor-microscale $\lambda$ is defined as $\lambda = \sqrt{15 \nu \sigma_x^2/\varepsilon}$ (note in this case the amplitude of the velocity fluctuations $\sigma_x$ correspond to those computed from the hot wire data) .

Figure~\ref{fig:HW}(b) shows the third-order longitudinal structure function (see Eq.~(\ref{eqS3})) normalised by the predicted Kolmogorov scaling $S_3(\ell) /(\varepsilon \ell)$ \citep{Frisch1995}. We also took hot-wire measurements at the same position for different mean flow freestream velocities, or equivalently, for $R_\lambda= 382$ and $R_\lambda = 507$. Their third-order structure function are also shown in panel (b) of Fig.~\ref{fig:HW} for comparison.
The amplitude of the normalised structure function seems to be highly sensitive to the value of the Reynolds number; at the largest $R_\lambda$ considered here its amplitude is above the prediction given by $S_3 = -4/5~\varepsilon \ell $ (note the dashed horizontal line in Fig.~\ref{fig:HW}(b) at $0.8$). This behaviour is consistent with previous studies discussed in the Introduction (see also \cite{zheng2021turbulent}), that report anomalous structure functions at $x/M=30$ and a normal behaviour at $x/M=80$. Our results also show that this effect is strongly dependent on the value of $R_\lambda$.

As for the case of static grids, anomalous behaviour seen in $S_3(\ell)$ has been linked to not fully developed turbulent flows \cite{zheng2021turbulent}. Given the non-Gaussian nature of the velocity PDFs seen here near the AG, we wonder if the origin of the anomaly in the scaling of the structure function in AG-generated turbulence could potentially be linked to the non-zero third-order moment observed for the velocity near the AG.

\begin{figure*}[t!]
\centering
\includegraphics[width=1.0\textwidth]{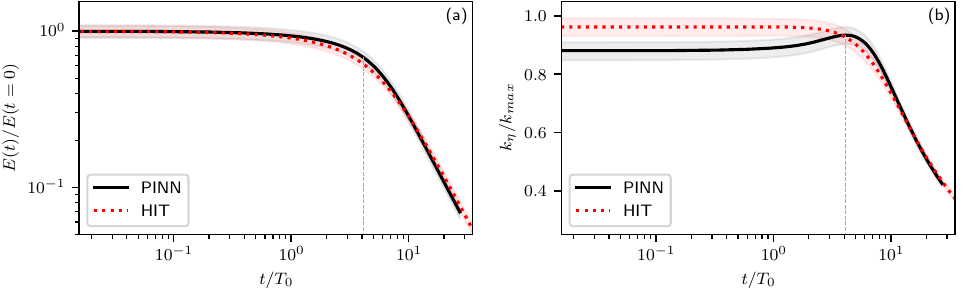}
\hfill
\includegraphics[width=1.0\textwidth]{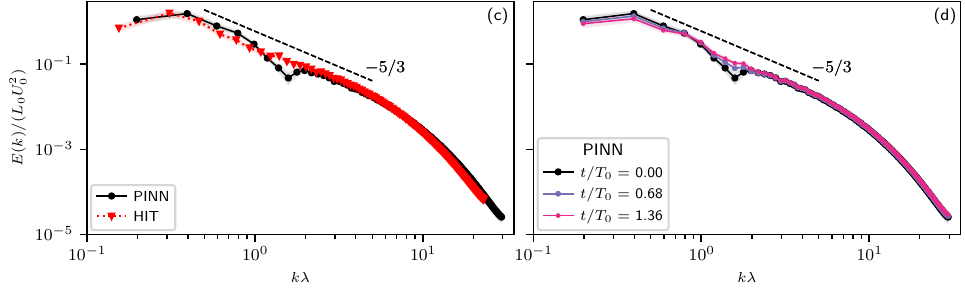}
\caption{
(a) Time decay of the kinetic energy for the PINN-generated states and for HIT simulations.  
(b) Time evolution of the Kolmogorov wavenumber, normalised by the largest wavenumber resolved by the simulations. The vertical dotted line indicates the time where $k_\eta$ peaks for the PINN-generated states, which is also indicated in panel (a).
(c) Initial kinetic energy spectrum $E(k)$ of the PINN-generated states and of the HIT simulations.
(d) Kinetic energy spectrum of the PINN-states at three successive times.
In all panels the thick solid and dotted lines, and the circular and triangular markers, correspond to the mean value over the 10 realisations, while the shaded bands represent typical variations between the different realisations.
}
\label{fig:decay}
\end{figure*}

\section{Decay of prepared states}

Turbulence generated in a wind tunnel can be studied numerically by means of DNSs. On the one hand, the flow at a fixed distance from the grid is (with certain limitations) comparable to simulations in a statistically steady state, in which energy is constantly supplied to the system by means of a forcing. On the other hand, one can compare the flow in the wind tunnel at different distances from the grid with simulations of a freely decaying flow (i.e., one in which no energy is input in the system as it evolves in time).
In order to numerically explore if initial conditions which present deviations from Gaussianity, as we observe in our experiments, result in an anomaly in the third-order structure function once turbulence has fully developed and has reached a self-similar decaying regime, we need to create specific initial conditions. This task would be ``trivial'' if we had access to the full 3D velocity field in the production zone (i.e., in the vicinity of the AG). However, we only have access to statistical information of the velocity field, as e.g., its centralised third-order moment.
 
\subsection{PINN and nudging protocol for initial conditions}

As discussed above, a combination of PINNs and the {\it nudging} data-assimilation technique will be used to prepare initial conditions for the DNSs compatible with the observations near the AG. We briefly recall the main points of the used method, which was introduced and validated using synthetic data in \citep{Angriman2023}.
A PINN is a neural network in which the loss term in the training is combined with physical information of a given system. For instance, a prediction may be penalised so that it satisfies a given physical law, or so that it is the solution to a given differential equation (such as the Navier-Stokes equation). In the implementation used in this work the PINN generates predictions which are compatible with an evolution given by the incompressible Navier-Stokes equations, and whose centralised third-order moment is compatible with the observed PDF asymmetry in the wind tunnel at $x=3M$. To do this we use a loss function
\begin{equation}
	L = L_d + \lambda_p L_p + \lambda_s L_s,
	\label{eq:loss_general}
\end{equation}
where
\begin{equation}
	L_d = \frac{1}{N_b} \sum\limits_{\{ i \}} (\mathbf{u}_i - \mathbf{u}_i^0)^2,
    \label{eq:loss_data}
\end{equation}
is the usual data term, and where $\mathbf{u}_i^0$ is an initial seed. Here the subindex $i$ labels the point and time at which the fields are evaluated, i.e., $\mathbf{u}_i = \mathbf{u}(x_i, y_i, z_i, t_i)$, and the summation is performed over $N_b$ different mini-batches $\{ i\}$. The parameters $\lambda_p$ and $\lambda_s$ are hyper-parameters that balance the importance of each term in the total loss function $L$. Also,
\begin{align}
L_p  =& \frac{1}{N_b}\sum_{\{i\}} \left[\left(
\frac{\partial \mathbf{u}_i}{\partial t}
+ \mathbf{u}_i\cdot \mathbf{\nabla} \mathbf{u}_i
+ \mathbf{\nabla} p_i - \nu \nabla^2 \mathbf{u}_i \right)^2
+ \left( \mathbf{\nabla} \cdot \mathbf{u}_i \right)^2 \right],
\end{align}
is the physics term, which forces $\mathbf{u}$ to be compatible with the Navier-Stokes equations and divergence-free, and where $p$ is the pressure per unit mass density. 
Finally, the term
\begin{align}
L_s  =  \;\; &\left(\frac{1}{N_b}\sum_{\{i\}} u_i \right)^2 
\\
\nonumber
+&\left[ \sqrt{\frac{1}{N_b}\sum_{\{i\}} u^2_i - \left(\frac{1}{N_b}\sum_{\{i\}} u_i \right)^2}  - \sigma_0 \right]^2
\\
\nonumber
+&\left[ {\frac{1}{N_b}\sum_{\{i\}} \left( u_i - \frac{1}{N_b}\sum_{\{i\}} u_i \right)^3}  - s_0^3 \right]^2 ,
\end{align}
is the loss function that takes care of imposing moments of the $x$-component of the field: the first term keeps the mean value at zero, and the second term fixes the standard deviation $\sigma_0$ (which is equivalent to setting the one-dimensional r.m.s.~velocity). The last term imposes the centralised third-order moment to be $s_0$ (note that it has dimensions of velocity). We would like for $s_0$ and $\sigma_0$ to be such that $s_0/\sigma_0 = s_{LDV}/\sigma_{x,LDV}$. In order to keep the velocity of order one we choose $\sigma_0 = 0.5 U_0$, and then $s_0 = \sigma_0 ~s_{LDV}/\sigma_{x,LDV} \approx 0.25 U_0$.
Note that while the loss terms $L_d$ and $L_p$ are applied to the three components of the velocity field, $L_s$ involves only the $x$ component of $\mathbf{u}$, $u$, so that we can interpret this direction as the streamwise direction in the force-free decay of the prepared turbulent states.
Details on the PINN architecture, choice of hyper-parameters, and training of the neural network can be found in \cite{Angriman2023}.

\begin{figure*}[t!]
\centering
\includegraphics[width=1.0\textwidth]{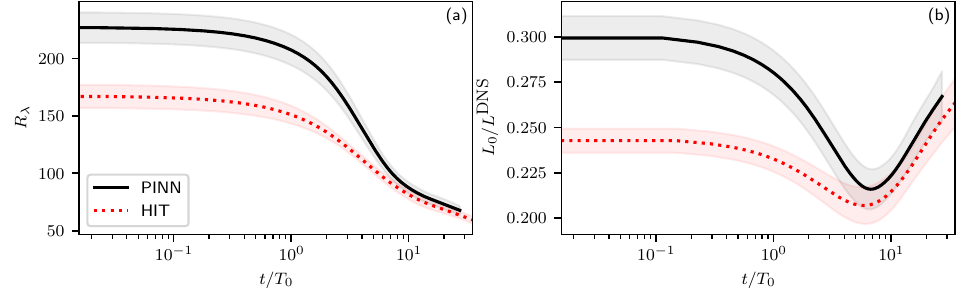}
\hfill
\caption{
(a) Evolution of the Taylor-scale Reynolds number. 
(b) Integral length scale $L_0$ as a function of time.
In both panels the thick solid and dotted lines correspond to the mean value over the 10 realisations, while the shaded bands represent typical variations between the different realisations; labels for panel (b) are the same as panel (a).}
\label{fig:decay_2}
\end{figure*}

Velocity fields using the PINN are generated as follows. An initial seed ${\bf u}^0$ for Eq.~(\ref{eq:loss_data}) is obtained from a low resolution DNS of ``HIT'' (using $32^3$ grid points). The gradient of Eq.~\eqref{eq:loss_general} is then evaluated iteratively, successively updating the weights of the neural network until the statistical moments of the output match the target experimental values (within fluctuations). We remark that, once the training is complete, the generated velocity field ${\bf u}({\bf x},t)$ has only $32^3$ grid points. This field is a divergence-free solution of the Navier-Stokes equation (within the errors of the PINN) with the imposed moments in the $u$ velocity field component.

We then sample the $32^3$ output of the PINN in the desired working grid resolution, which is $512^3$ grid points in our case. However, the PINN states do not contain information compatible with the Navier-Stokes equations at small scales. In order to obtain higher Reynolds number states, closer to those achievable in the experiment, we use the nudging technique. Briefly, this method evolves the equations of motion of the system with an additional relaxation term that penalises the field $\mathbf{u}$ when it strays away from a given reference field $\mathbf{u}_\text{ref}$. In our case, $\mathbf{u}_\text{ref}$ corresponds to the velocity field sampled from the final state of the PINN. Hence, the equations solved in the nudging stage using a DNS code are
\begin{equation}
    \partial_t {\mathbf u} + ({\mathbf u}\cdot {\mathbf \nabla}) {\mathbf u} = -{\mathbf \nabla} p + \nu \nabla^2 {\mathbf u} - \alpha \altmathcal{I}({\mathbf u} - {\mathbf u}_\text{ref}),
\label{eq:nudging}
\end{equation}
where the last term on the right hand side corresponds to the nudging term, which penalizes the distance between the reference data ${\mathbf u}_\text{ref}$ and $\mathbf{u}$. The amplitude of this term is controlled by $\alpha$, and $\altmathcal{I}$ is a filter that acts only where the data is available. This filter can be applied in real or in Fourier space. In particular, we use a low-pass filter in Fourier space, which projects the spatial part of $\mathbf{u}$ on the Fourier modes with normalised wave number $k \in [k_0=0, k_1=9]$, the modes in which ${\mathbf u}_\text{ref}$ contains the most relevant spectral information. Thus, the effect of the filter is expressed as
\begin{equation}
    \altmathcal{I}{\mathbf u}({\mathbf x}, t) = \sum\limits_{k_0 \leq \lvert{\mathbf k}\lvert \leq k_1} \hat{\mathbf{u}}(\mathbf{k},t)~ \text{exp}(i{\mathbf k}\cdot {\mathbf x}).
\end{equation}

Evolution of Eq.~\eqref{eq:nudging} results in a new velocity field ${\mathbf u}$ that has a broad inertial range (i.e., it is turbulent) and is compatible at large scales with the observed anomalies in the moments of $u$ (i.e., it is close to ${\mathbf u}_\text{ref}$ at large scales). For the evolution, Eq.~\eqref{eq:nudging} is written in dimensionless units based on a unit length $L^\text{DNS}$ and a unit velocity $U^\text{DNS}$, and solved using the parallel pseudospectral code GHOST \citep{Mininni2011, Rosenberg2020, Fontana_2020}. The solving domain corresponds to a three-dimensional box of side $2\pi L^\text{DNS}$ with periodic boundary conditions, with the initial condition given by $\mathbf{u}(t=0) = \mathbf{u}_\text{ref}(t=0)$. Time integration is done for the time interval for which reference data is available, i.e., the temporal window in which the neural network was trained. 
As previously mentioned, after the nudging protocol is applied, the obtained fields have information at small scales that is compatible with a turbulent flow, while retaining statistical moments imposed by the neural network.

\subsection{Free decay of prepared states}

Having the tools to generate high resolution velocity fields with statistical moments that resemble those observed in the flow in the vicinity of the active grid, we study how these fields evolve in time without any energy input. The goal is to see if the statistical anomalies observed near the active grid in the experiments can cause the anomalies that are later observed in the decay in the experiment.
We consider $10$ different realisations of the ``PINN+nudging'' protocol, changing the initial seed that feeds the neural network in each case to obtain different states. From each realisation we take the velocity field at a given time, and we use it as the initial condition of a $512^3$ DNS, in which the evolution is given by the force-free incompressible Navier-Stokes equations (i.e., Eq.~\eqref{eq:nudging} without the nudging term).
By doing this we can study the free decay of the states, and compare it with the flow generated throughout the test section in the wind tunnel; we will label these flows and datasets with the superscript ``PINN.'' As a reference, we also consider the free decay of initial conditions corresponding to HIT. In order to prepare them, we first evolve the equations of motion with a random forcing to sustain the turbulence, using a resolution of $512^3$ grid points. The forcing injects energy in the Fourier shell $k L^\text{DNS} \in [1,3]$ with fixed amplitude, and slowly-varying phases with a correlation time of $0.05~ L^\text{DNS}/U^{DNS}$ . The system is evolved until a steady state is reached, and afterwards the forcing is turned off and the flow is left to decay freely. An ensemble of 10 realisations is also used for the HIT states.

\begin{figure*}[t!]
\centering
\includegraphics[width=1.0\textwidth]{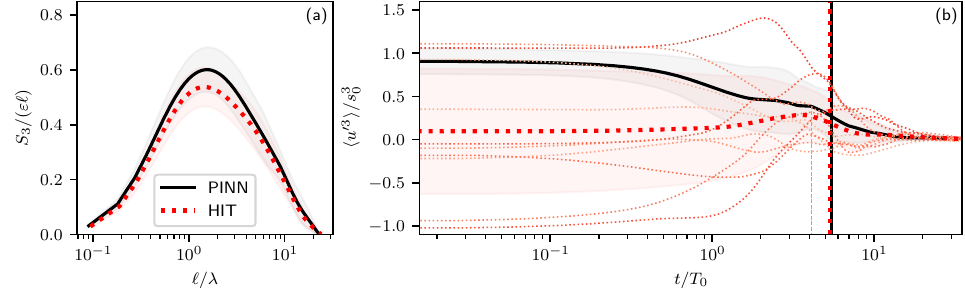}
\hfill
\includegraphics[width=1.0\textwidth]{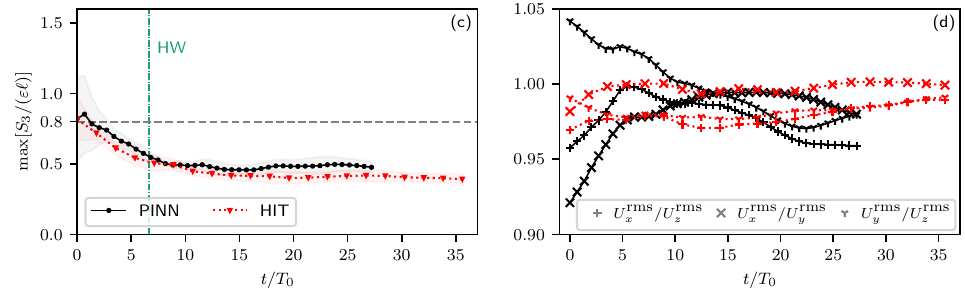}
\caption{(a) Third-order longitudinal structure function, compensated by $\varepsilon \ell$, for the HIT runs and for the PINN-generated states, at time $t/T_0 \approx 5$. The thick solid and dotted lines correspond to the mean value over the 10 realisations, while the shaded bands represent typical variations between the different realisations. (b) Evolution of the third-order moment of the streamwise component of the velocity, normalised by the value imposed during the PINN protocol. Thick and shaded regions are as in panel (a). The thin dotted lines correspond to the time evolution in each one of the single HIT simulations. The black solid and red dotted vertical lines indicate the time at which the structure functions in panel (a) are computed. The dashed vertical line indicated the time where $xk_\eta$ peaks for the PINN-generated states (see Fig.~\ref{fig:decay}). (c) Peak value of the normalised third-order structure function along the decay. The dash-dotted vertical line indicates the equivalent time at which the hot-wire measurements in the wind tunnel are performed. (d) Ratio of r.m.s.~values of the fluctuating velocity components, shown in black and red markers for PINN-generated and HIT states, respectively.}
\label{fig:S3}
\end{figure*}

For both the free decay of PINN and HIT states, the simulations are performed using the GHOST code, under similar conditions as the ones corresponding to the nudging simulations. We use kinematic viscosities $\nu^\text{PINN} = 5 \times 10^{-4} L^\text{DNS} U^\text{DNS}$ and $\nu^\text{HIT} = 5.5 \times 10^{-4} L^\text{DNS} U^\text{DNS}$. All simulations are such that $k_\text{max}/k_\eta > 1$, where $k_\text{max} = N/3$ corresponds to the largest wave number resolved by the simulations, and $k_\eta = (\varepsilon/\nu^3)^{1/4}$ is the Kolmogorov wave number.

Figure~\ref{fig:decay}(a) shows the kinetic energy evolution for the PINN-prepared and the HIT states, normalised by the energy at $t=0$. The ensemble averages over the $10$ realisations are shown in solid and dashed lines, for the PINN-generated and the HIT states, respectively. The shaded areas indicate $\pm 1$ standard deviation between the realisations. Time is normalised by $T_0 = L_0/U_0$, the integral time scale estimated from inlet conditions. Here, 
\begin{equation}
	L_0 = \frac{\pi}{4}\frac{\int E(k)/k~dk}{\int E(k)~dk}
	\label{eq:Lint_DNS}
\end{equation}
is the Eulerian integral scale computed from the 1D energy spectrum corresponding to the streamwise component of the velocity field at $t=0$. $U_0$ is the r.m.s.~value of $u$, also at the start of the evolution. From $t/T_0 \approx 4$ both flows display a self-similar decay. Disregarding a time offset, it seems that the decay exponent for the PINN-states is larger than for the HIT-states. However, when considering a dependence of the type
\begin{equation}
	E = E_0~ (t-t_0)^\alpha,
\end{equation}
imposing $\alpha^\text{HIT} = \alpha^\text{PINN} \equiv \alpha$, results in non-zero values of $t_0^\text{HIT}/T_0^\text{HIT} = -0.53$ and $t_0^\text{PINN}/T_0^\text{PINN}=0.19$, and yields a value of $\alpha = -1.40$ compatible with the literature (that also reports slightly smaller values, between $-1.1$ and $-1.3$ \cite{thormann2014decay,sinhuber2015decay}, specially for experiments \cite{Panick_2022}).

Figure~\ref{fig:decay}(b) shows the Kolmogorov wave number throughout the evolution. While for the HIT simulations $k_\eta$ remains constant until the beginning of the self-similar decay, interestingly we observe an increase of $k_\eta$ for the PINN-prepared states at around the time when the self-similar decay starts, which systematically occurs in all of the realisations. The vertical dashed lines in panels (a) and (b) of Fig.~\ref{fig:decay} indicate the time when the maximum value of $k_\eta$ is realised. This increase in $k_\eta$ implies that the range of scales involved in the energy cascade is increasing, meaning that the production of turbulence in the PINN-generated states differs from the production in the HIT case, although we do not observe that this has an impact on the exponent of the self-similar decay that takes place afterwards. Note that this increase in $k_\eta$ takes place even after preparing the initial PINN flows with the nudging technique that results in a broad inertial range, suggesting that the anomaly in the third-order moment at large-scales in these states results in further turbulence production even after turning off the forcing.

We can gain further insight in this respect by comparing the kinetic energy spectrum $E(k)$ of the PINN-generated states and of the HIT simulations. The average spectra over all realisations at $t=0$ for both flow types are shown in Fig.~\ref{fig:decay}(c). Both flows display more than one decade of power-law like behaviour with an exponent close to $-5/3$, compatible with fully-developed turbulent conditions. The PINN spectrum displays a slight dip close to $k \lambda \approx 1.5$, which stems from the upper bound of the Fourier band-pass filter used in the nudging stage (Eq.~\eqref{eq:nudging}). This dip quickly disappears as the flow freely decays, as it can be seen in panel (d) in Fig.~\ref{fig:decay}, where we show three successive $E(k)$ curves. At $t/T_0 = 0.68$ the amplitude of the dip has already decreased significantly, and at $t/T_0 = 1.36$ the PINN spectrum is almost indistinguishable from that of HIT. Note that this time corresponds to the early stages of the decay, before the peak of $k_\eta$ occurs (indicated in panel (b) in Fig.~\ref{fig:decay}). Hence, throughout the most part of the evolution, the PINN and HIT states are similar from a spectral viewpoint.

The evolution of the Taylor-based Reynolds number $R_\lambda$ is shown in Fig.~\ref{fig:decay_2}(a). Even when initially $R_\lambda$ is larger for the PINN-states than for HIT, at around $t/T_0 \approx 7$ the values of $R_\lambda$ of the two types of flows converge. In Fig.~\ref{fig:decay_2}(b) we plot the integral length scale $L_0(t)$, computed as in Eq.~\eqref{eq:Lint_DNS} with the corresponding spectrum $E(k,t)$. As with active-grids, which are characterised by a larger integral length scale than the regular, passive grids (as the former are in the order of the tunnel lateral size and the latter of the mesh size $M$), the PINN-generated states also present a higher value of $L_0$ when compared with the HIT states. After $k_\eta$ has reached its peak, the integral length scale has a dip and then the values overlap with the those of HIT.

We now move on to analyse what the third-order longitudinal structure function looks like in the simulations, at a time when the turbulence is already developed. In analogy with the hot-wire experiments, we compute $S_3(\ell)$ for the $x$-component of the velocity field. For the HIT fields all three components are statistically equivalent, but in the PINN-generated states this is the component whose third-order moment was imposed. Figure~\ref{fig:S3}(a) shows $S_3(\ell)$ normalised by $\varepsilon \ell$, at $t/T_0 \approx 5$. As before, the solid and dashed lines correspond to the average over the 10 realisations of each flow, and the shaded bands represent the typical variation. We observe that the initial conditions generated with the PINN systematically display larger amplitudes than the HIT states. Even though this amplitude is not as high as the one observed in the wind tunnel (note that the DNSs have a smaller $R_\lambda$ than the experiments), it represents a clear deviation from the HIT cases.

Figure~\ref{fig:S3}(b) shows the evolution of the centralised third-order moment of $u$, normalised by the target value of the neural network $s_0$, for the PINN and HIT states. The solid and dotted vertical lines indicate the times when the structure functions shown in Fig.~\ref{fig:S3}(a) were computed. We also plot the individual evolution of each HIT realisation, with a finer line width. 

As the centralised third-order moment of $u$ is imposed at $t=0$  in the PINN-states, $\langle u'^3 \rangle/s_0^3$ displays less dispersion in those runs than in the HIT runs (confirmed by the shaded grey area being narrower than the red shaded region). In spite of this large dispersion in the HIT runs, with some runs with large values of $\langle u'^3 \rangle/s_0^3$, the HIT ensemble averages to $\langle u'^3 \rangle/s_0^3 \approx 0$, while the PINN-states have $\langle u'^3 \rangle/s_0^3 > 0$ and systematically display an anomaly in this quantity for long times. This suggests that the observed anomalies near the active grid, which are imposed by the PINN in the DNSs, introduce correlations in the flow that result in it being statistically different from an idealised HIT flow, at least in the initial part of the free decay.

This is further supported by considering the time evolution of the maximum value of $S_3/(\varepsilon \ell)$, shown in Fig.~\ref{fig:S3}(c). We also indicate with a vertical line the time $t/T_0$ that corresponds to the hot-wire measurements performed in the wind tunnel (shown in Fig.~\ref{fig:HW}). While for the HIT simulations the peak value quickly drops below $4/5$, the peak for the PINN runs is, on average, larger than the expected value for longer periods of time. Even though the experimental structure function remains anomalous for larger times than in the simulations (i.e., the peak value of $S_3/\varepsilon\ell$ is larger than $4/5$ for longer times), imposing the initial centralised third-order moment in turbulent states has an effect on the evolution of the structure function that qualitatively shifts the HIT behaviour towards something closer to the experiment. We note however that $R_\lambda$ in simulations is smaller than in experiments. As we have shown for the experiments in Fig.~\ref{fig:HW}(b), the anomalous behaviour in the amplitude of $S_3$ depends on the value of $R_\lambda$, so it would be interesting to evaluate how an increase in the Reynolds number affects the numerical results (a detailed study of Reynolds effects is out of the scope of this work, see also \cite{Antonia2019}). However, it should be kept in mind that the differences in the simulations with respect to the experiment is not just in the value of $R_\lambda$, but also in the fact that we only impose the centralised third order moment of $u'$. In the experiments, higher order moments of $u'$, as well as cross-correlations between the different velocity components, may also be partially responsible for the anomalous behaviour observed downstream. Finally, when looking at the ratio of r.m.s.~component-wise velocities, shown in Fig.~\ref{fig:S3}(d), the PINN-generated states present acceptable levels of isotropy which are comparable with experimental observations in the same wind tunnel \citep{Mora2019}.

\section{Conclusions}

Wind tunnels are an essential tool in the study of environmental flows, providing a controlled setup to replicate in the laboratory the complex interactions between wind and various structures. In the design of aircraft, wind turbines, and wind farms, wind tunnels allow engineers to optimise wing and blade shapes and configurations by testing different models under varying wind conditions which include turbulent scenarios. Additionally, wind tunnels are invaluable for studying idealised (i.e., homogeneous and isotropic) turbulence, specially in recent years with the introduction of active grids that allowed generation of flows with very large Reynolds numbers \cite{Mydlarski2017}. When combined with numerical simulations, they provide data that is crucial for understanding and predicting turbulent phenomena \cite{Panick_2022}.

However, anomalies observed in the near field of the grid in wind tunnel experiments \cite{zheng2021turbulent}, and difficulties in reproducing these flows in simulations when only partial statistical information is available from observations \cite{Angriman2023}, have raised questions on the properties of the turbulent flow generated by active grids when using certain protocols. We have presented a case study on how to combine data from real experiments with machine learning, data assimilation, and numerical simulations, to shed light on some of these questions.

Laboratory measurements of the flow near the active grid using Laser Doppler Velocimetry indicate the existence of anomalies in the fluid velocity, including flow reversals, and in particular, the existence of systematic asymmetries in the statistics of the streamwise fluctuating velocity component. This anomaly was quantified in our study using the centralised third-order moment of the velocity. Generation of 
compatible third-order moment anomalies in initial conditions for numerical simulations, using PINNS and a nudging data assimilation method, indicate that these anomalies persist for long times, and can give rise downstream to larger values of the longitudinal third-order structure function than those expected for homogeneous and isotropic turbulence.

Several studies report that far from the grid (for this flow, 70 meshes downstream or more), two-point statistics, isotropy and homogeneity reach universal behaviour~ \cite{Kuchler2023,mydlarski1996onset,ertuncc2007experimental}. Considering this distance, anomalies detected in the active grid thus persist significantly farther downstream when compared to those present in static grid flows. Our results confirm that, depending on the flow Reynolds number, anomalies are still present as far as 40 grid meshes downstream. This is a range relevant for several wind tunnel studies, and future studies and applications should have particular care when analysing these flows. Moreover, the methodology presented here can also be used to characterize these flows, and to find ways to ameliorate or control these effects, as active grids have a central role when trying to reach large Reynolds numbers in experiments.

The protocol presented here for data assimilation can be further extended for other applications in mechanical and aerospace engineering, in which only incomplete or statistical information of the flow is available to prepare initial conditions for numerical simulations.


\section*{Acknowledgements}
The authors thank Amélie Ferran for providing the hot wire measurements. 
The authors acknowledge funding by ECOS-Sud project No.~A18ST04. PCDL, PJC, and PDM financial support from UBACyT Grant No.~20020170100508BA and Redes Federales de Alto Impacto, Argentina.

\bibliographystyle{elsarticle-num} 
\bibliography{main}

\begin{thebibliography}{10}
\expandafter\ifx\csname url\endcsname\relax
  \def\url#1{\texttt{#1}}\fi
\expandafter\ifx\csname urlprefix\endcsname\relax\def\urlprefix{URL }\fi
\expandafter\ifx\csname href\endcsname\relax
  \def\href#1#2{#2} \def\path#1{#1}\fi

\bibitem{makita1991active}
H.~Makita, K.~Sassa, Active turbulence generation in a laboratory wind tunnel,
  in: Advances in Turbulence 3: Proceedings of the Third European Turbulence
  Conference Stockholm, July 3--6, 1990, Springer, 1991, pp. 497--505.

\bibitem{Mydlarski2017}
L.~Mydlarski, A turbulent quarter century of active grids: from {M}akita (1991)
  to the present, Fluid Dynamics Research 49~(6) (2017) 061401.
\newblock \href {https://doi.org/https://doi.org/10.1088/1873-7005/aa7786}
  {\path{doi:https://doi.org/10.1088/1873-7005/aa7786}}.

\bibitem{Hearst2015}
R.~J. Hearst, P.~Lavoie, The effect of active grid initial conditions on high
  reynolds number turbulence, Experiments in Fluids 56~(10) (2015) 185.
\newblock \href {https://doi.org/10.1007/s00348-015-2052-1}
  {\path{doi:10.1007/s00348-015-2052-1}}.

\bibitem{hearst2017tailoring}
R.~J. Hearst, B.~Ganapathisubramani, Tailoring incoming shear and turbulence
  profiles for lab-scale wind turbines, Wind Energy 20~(12) (2017) 2021--2035.

\bibitem{neuhaus2021exploring}
L.~Neuhaus, F.~Berger, J.~Peinke, M.~H{\"o}lling, Exploring the capabilities of
  active grids, Experiments in Fluids 62~(6) (2021) 130.

\bibitem{wester2022design}
T.~T. Wester, J.~Krauss, L.~Neuhaus, A.~H{\"o}lling, G.~G{\"u}lker,
  M.~H{\"o}lling, J.~Peinke, How to design a 2d active grid for dynamic inflow
  modulation, Flow, Turbulence and Combustion 108~(4) (2022) 955--972.

\bibitem{scott2020wind}
R.~Scott, B.~Viggiano, T.~Dib, N.~Ali, M.~H{\"o}lling, J.~Peinke, R.~B. Cal,
  Wind turbine partial wake merging description and quantification, Wind Energy
  23~(7) (2020) 1610--1618.

\bibitem{hamilton2012statistical}
N.~Hamilton, H.~Suk~Kang, C.~Meneveau, R.~Bayo{\'a}n~Cal, Statistical analysis
  of kinetic energy entrainment in a model wind turbine array boundary layer,
  Journal of Renewable and Sustainable Energy 4~(6) (2012).

\bibitem{gambuzza2021effects}
S.~Gambuzza, B.~Ganapathisubramani, The effects of free-stream turbulence on
  the performance of a model wind turbine, Journal of Renewable and Sustainable
  Energy 13~(2) (2021).

\bibitem{larssen2011generation}
J.~V. Larssen, W.~J. Devenport, On the generation of large-scale homogeneous
  turbulence, Experiments in Fluids 50 (2011) 1207--1223.

\bibitem{bodenschatz2014variable}
E.~Bodenschatz, G.~P. Bewley, H.~Nobach, M.~Sinhuber, H.~Xu, Variable density
  turbulence tunnel facility, Review of Scientific Instruments 85~(9) (2014).

\bibitem{sinhuber2017dissipative}
M.~Sinhuber, G.~P. Bewley, E.~Bodenschatz, Dissipative effects on
  inertial-range statistics at high reynolds numbers, Physical review letters
  119~(13) (2017) 134502.

\bibitem{mydlarski1996onset}
L.~Mydlarski, Z.~Warhaft, On the onset of high-reynolds-number grid-generated
  wind tunnel turbulence, Journal of Fluid Mechanics 320 (1996) 331--368.

\bibitem{Kuchler2023}
C.~K{\"u}chler, G.~P. Bewley, E.~Bodenschatz, Universal velocity statistics in
  decaying turbulence, Physical Review Letters 131~(2) (2023) 024001.
\newblock \href
  {https://doi.org/https://doi.org/10.1103/PhysRevLett.131.024001}
  {\path{doi:https://doi.org/10.1103/PhysRevLett.131.024001}}.

\bibitem{Mora2019}
D.~Mora, E.~M. Pladellorens, P.~R. Turr{\'o}, M.~Lagauzere, M.~Obligado, Energy
  cascades in active-grid-generated turbulent flows, Physical Review Fluids
  4~(10) (2019) 104601.
\newblock \href
  {https://doi.org/https://doi.org/10.1103/PhysRevFluids.4.104601}
  {\path{doi:https://doi.org/10.1103/PhysRevFluids.4.104601}}.

\bibitem{sinhuber2015decay}
M.~Sinhuber, E.~Bodenschatz, G.~P. Bewley, Decay of turbulence at high reynolds
  numbers, Physical Review Letters 114~(3) (2015) 034501.

\bibitem{zheng2023unsteady}
Y.~Zheng, K.~Nakamura, K.~Nagata, T.~Watanabe, Unsteady dissipation scaling in
  static-and active-grid turbulence, Journal of Fluid Mechanics 956 (2023) A20.

\bibitem{Obligado2014}
M.~Obligado, T.~Teitelbaum, A.~Cartellier, P.~D. Mininni, M.~Bourgoin,
  Preferential concentration of heavy particles in turbulence, Journal of
  Turbulence 15~(5) (2014) 293--310.
\newblock \href {https://doi.org/https://doi.org/10.1080/14685248.2014.897710}
  {\path{doi:https://doi.org/10.1080/14685248.2014.897710}}.

\bibitem{sumbekova2017preferential}
S.~Sumbekova, A.~Cartellier, A.~Aliseda, M.~Bourgoin, Preferential
  concentration of inertial sub-kolmogorov particles: the roles of mass loading
  of particles, stokes numbers, and reynolds numbers, Physical Review Fluids
  2~(2) (2017) 024302.

\bibitem{prakash2016energy}
V.~N. Prakash, J.~M. Mercado, L.~Van~Wijngaarden, E.~Mancilla, Y.~Tagawa,
  D.~Lohse, C.~Sun, Energy spectra in turbulent bubbly flows, Journal of Fluid
  Mechanics 791 (2016) 174--190.

\bibitem{ferran2023experimental}
A.~Ferran, N.~Machicoane, A.~Aliseda, M.~Obligado, An experimental study on the
  settling velocity of inertial particles in different homogeneous isotropic
  turbulent flows, Journal of Fluid Mechanics 970 (2023) A23.

\bibitem{kang2003decaying}
H.~S. Kang, S.~Chester, C.~Meneveau, Decaying turbulence in an
  active-grid-generated flow and comparisons with large-eddy simulation,
  Journal of Fluid Mechanics 480 (2003) 129--160.

\bibitem{Frisch1995}
U.~Frisch, Turbulence: the legacy of AN Kolmogorov, Cambridge university press,
  1995.
\newblock \href {https://doi.org/https://doi.org/10.1017/CBO9781139170666}
  {\path{doi:https://doi.org/10.1017/CBO9781139170666}}.

\bibitem{zheng2021turbulent}
Y.~Zheng, K.~Nagata, T.~Watanabe, Turbulent characteristics and energy transfer
  in the far field of active-grid turbulence, Physics of Fluids 33~(11) (2021).

\bibitem{kuchler2021measurements}
C.~K{\"u}chler, Measurements of turbulence at high reynolds numbers: From
  eulerian statistics towards lagrangian particle tracking, Ph.D. thesis,
  Dissertation, G{\"o}ttingen, Georg-August Universit{\"a}t, 2020 (2021).

\bibitem{batchelor1948decay}
G.~K. Batchelor, A.~A. Townsend, Decay of isotropic turbulence in the initial
  period, Proceedings of the Royal Society of London. Series A. Mathematical
  and Physical Sciences 193~(1035) (1948) 539--558.

\bibitem{krogstad2010grid}
P.-{\AA}. Krogstad, P.~Davidson, Is grid turbulence saffman turbulence?,
  Journal of Fluid Mechanics 642 (2010) 373--394.

\bibitem{larssen2005large}
J.~V. Larssen, Large scale homogeneous turbulence and interactions with a
  flat-plate cascade, Ph.D. thesis, Virginia Polytechnic Institute and State
  University (2005).

\bibitem{puga2017normalized}
A.~J. Puga, J.~C. LaRue, Normalized dissipation rate in a moderate taylor
  reynolds number flow, Journal of Fluid Mechanics 818 (2017) 184--204.

\bibitem{mora2020estimating}
D.~O. Mora, M.~Obligado, Estimating the integral length scale on turbulent
  flows from the zero crossings of the longitudinal velocity fluctuation,
  Experiments in Fluids 61~(9) (2020) 199.

\bibitem{shet2017eulerian}
C.~S. Shet, M.~R. Cholemari, S.~V. Veeravalli, Eulerian spatial and temporal
  autocorrelations: assessment of taylor's hypothesis and a model, Journal of
  Turbulence 18~(12) (2017) 1105--1119.

\bibitem{Raissi2019}
M.~Raissi, P.~Perdikaris, G.~E. Karniadakis, Physics-informed neural networks:
  {A} deep learning framework for solving forward and inverse problems
  involving nonlinear partial differential equations, Journal of Computational
  physics 378 (2019) 686--707.
\newblock \href {https://doi.org/https://doi.org/10.1016/j.jcp.2018.10.045}
  {\path{doi:https://doi.org/10.1016/j.jcp.2018.10.045}}.

\bibitem{Clark2018}
P.~Clark Di~Leoni, A.~Mazzino, L.~Biferale, Inferring flow parameters and
  turbulent configuration with physics-informed data assimilation and spectral
  nudging, Physical Review Fluids 3~(10) (2018) 104604.
\newblock \href
  {https://doi.org/https://doi.org/10.1103/PhysRevFluids.3.104604}
  {\path{doi:https://doi.org/10.1103/PhysRevFluids.3.104604}}.

\bibitem{Clark2020}
P.~Clark Di~Leoni, A.~Mazzino, L.~Biferale, Synchronization to big data:
  {N}udging the {N}avier-{S}tokes equations for data assimilation of turbulent
  flows, Physical Review X 10~(1) (2020) 011023.
\newblock \href {https://doi.org/https://doi.org/10.1103/PhysRevX.10.011023}
  {\path{doi:https://doi.org/10.1103/PhysRevX.10.011023}}.

\bibitem{Angriman2023}
S.~Angriman, P.~Cobelli, P.~D. Mininni, M.~Obligado, P.~Clark Di~Leoni,
  Assimilation of statistical data into turbulent flows using physics-informed
  neural networks, The European Physical Journal E 46~(3) (2023) 13.
\newblock \href
  {https://doi.org/https://doi.org/10.1140/epje/s10189-023-00268-9}
  {\path{doi:https://doi.org/10.1140/epje/s10189-023-00268-9}}.

\bibitem{Mininni2011}
P.~D. Mininni, D.~Rosenberg, R.~Reddy, A.~Pouquet, A hybrid {MPI}--open{MP}
  scheme for scalable parallel pseudospectral computations for fluid
  turbulence, Parallel computing 37~(6-7) (2011) 316--326.
\newblock \href {https://doi.org/https://doi.org/10.1016/j.parco.2011.05.004}
  {\path{doi:https://doi.org/10.1016/j.parco.2011.05.004}}.

\bibitem{Rosenberg2020}
D.~Rosenberg, P.~D. Mininni, R.~Reddy, A.~Pouquet, {GPU} parallelization of a
  hybrid pseudospectral geophysical turbulence framework using {CUDA},
  Atmosphere 11~(2) (2020) 178.
\newblock \href {https://doi.org/https://doi.org/10.3390/atmos11020178}
  {\path{doi:https://doi.org/10.3390/atmos11020178}}.

\bibitem{Fontana_2020}
M.~Fontana, O.~P. Bruno, P.~D. Mininni, P.~Dmitruk, Fourier continuation method
  for incompressible fluids with boundaries, Computer Physics Communications
  256 (2020) 107482.

\bibitem{thormann2014decay}
A.~Thormann, C.~Meneveau, Decay of homogeneous, nearly isotropic turbulence
  behind active fractal grids, Physics of Fluids 26~(2) (2014).

\bibitem{Panick_2022}
J.~Panickacheril~John, D.~A. Donzis, K.~R. Sreenivasan, Laws of turbulence
  decay from direct numerical simulations, Philosophical Transactions of the
  Royal Society A: Mathematical, Physical and Engineering Sciences 380 (2022)
  20210089.
\newblock \href {https://doi.org/10.1098/rsta.2021.0089}
  {\path{doi:10.1098/rsta.2021.0089}}.

\bibitem{Antonia2019}
R.~Antonia, S.~Tang, L.~Djenidi, Y.~Zhou, Finite {R}eynolds number effect and
  the 4/5 law, Physical Review Fluids 4~(8) (2019) 084602.
\newblock \href
  {https://doi.org/https://doi.org/10.1103/PhysRevFluids.4.084602}
  {\path{doi:https://doi.org/10.1103/PhysRevFluids.4.084602}}.

\bibitem{ertuncc2007experimental}
{\"O}.~Ertun{\c{c}}, Experimental and numerical investigations of axisymmetric
  turbulence, Ph.D. thesis, Erlangen, N{\"u}rnberg, Univ., Diss., 2006 (2007).

\end{thebibliography}

\end{document}